\def\la{\langle}
\def\ra{\rangle}
\def\vn{{\bf n}}
\def\gn{{\mathcal G}({\bf n})}
\def\tracce{{\hbox{Tr}}}
\def\haf{\frac12}
\def\third{\frac13}

\documentclass{article}  

\begin{document}

\begin{center}
{{\bf  Nonrelativisitic Ideal Gasses and Lorentz Violations\\}
\vglue 1.0cm
{Don Colladay and Patrick McDonald\\} 
\bigskip
{\it New College of Florida\\}
\medskip
{\it Sarasota, FL, 34243, U.S.A.\\}
 }
\end{center}
\vglue 0.8cm

\vglue 0.3cm
 
{\rightskip=3pc\leftskip=3pc\noindent
We develop statistical mechanics for a nonrelativisitic ideal gas in
the presence of Lorentz violating background fields.  The analysis is
performed using the Standard-Model Extension (SME).  We derive the
corresponding laws of thermodynamics and find that, to lowest order in
Lorentz violation, the scalar thermodynamic variables are corrected by
a rotationally invariant combination of the Lorentz terms which can be
interpreted in terms of a (frame dependent) effective mass.  We find
that spin couplings can induce a temperature independent polarization
in the gas that is not present in the conventional case.
}

\vskip 1 cm

\vskip 1 cm

\section{Introduction}
The Standard Model Extension (SME) provides a convenient 
framework for studying the effects of spontaneous Lorentz and CPT
symmetry breaking within the context of conventional quantum field
theory.\cite{ck}$^,$\cite{cpt98}  In this report we develop a
statistical mechanics formalism for calculations involving the SME.
Our approach and our results are quite general and include a complete
analysis of the effects of all Lorentz violating terms on a
nonrelativistic ideal gas. Complete details of the results announced
here will appear elsewhere.\cite{cm3}  
\section{Notation and Framework}
We adopt the viewpoint of Jaynes.\cite{jaynes}  In this approach to
statistical inference, one assumes a collection of states,
$\{\psi_i\}_{i=1}^\infty,$ and a finite collection of real valued
functions on the collection of states, $\{f_j\}_{j=1}^l.$  Given a
distribution of states, $q_i =   q(\psi_i),$ we  denote by brackets
the corresponding expectations; $\la  f_j \ra = \sum_i f_j(\psi_i)
q_i.$  Given observations of the mean values $\{\la f_j
\ra\}_{j=1}^l,$ Jaynes argues that the most likely distribution for
the given data is obtained by maximizing the (information) entropy, 
$S = -k\sum_i q_i \ln{(q_i)},$ subject to the constraints given by the
observations (here $k$ is a positive constant).  A formal argument via
variational calculus then leads immediately to a solution of the form 
\begin{eqnarray}
q_i & = & \frac{e^{-\sum_{j=1}^l \lambda_j f_j(\psi_i)}}{Z} 
\quad ,
\label{dist1}
\end{eqnarray}
where the Lagrange multipliers, $\lambda_j,$ are real constants and $Z$
is the partition function, $Z({\bf \lambda}) = \sum_{i} e^{-\sum_{j=1}^l
\lambda_j f_j(\psi_i)}.$  The same formalism permits the observables
$f_j$ to depend on a finite number of parameters.

When this formalism is applied to study the statistical mechanics of
an ideal gas, one immediately identifies $k$ with the Boltzmann
constant and the Lagrange multipliers with the usual thermodynamic
quantities (ie scaled inverse temperature, chemical potential, etc).
The central features of thermodynamics become consequences of
formal computation.  As was emphasized by Jaynes, the method provides
accurate thermodynamic properties of a system assuming empirically
accurate observations of the mean values and the correct laws of
motion embedded in the hamiltonian.  

To use this framework to study ideal gasses with Lorentz violation 
all that is required is the appropriate hamiltonian.  For
nonrelativistic spin-$\frac12$ fermions of mass $m$ this has been
worked out by Kostelecky and Lane.\cite{klane}  The result to second
order in $\frac{p}{m},$ $p$ momentum, is 
\begin{eqnarray}
H & =& \frac{p^2}{2m} + A + B_j\sigma^j + C_{j}\frac{p_j}{m} +
D_{jk}\frac{p_j}{m} \sigma^k + \nonumber \\ 
 &  &  F_{jk}\frac{p_jp_k}{2m} +
G_{jkl}\frac{p_jp_k}{2m} \sigma^l \quad , 
\label{H2}
\end{eqnarray}
where $A, \ B_j, \ C_j, \ D_{jk}, \ F_{jk}$ and $G_{jkl}$ are real
parameters which can be given explicitly in terms of the standard
collection of parameters defining the Lagrangian of the minimal
SME.\cite{klane}$^{,}$\cite{cm3}
\section{Single particle systems}
We first consider a system consisting of a single free spin-$\frac12$ 
particle governed by the hamiltonian $H$ appearing in (\ref{H2}), constrained
to a cube of side length $L$.  

Denote by $\psi^{(0)}_{\vn,s}$ the standard unperturbed solutions for
the hamiltonian where $\vn = (n_1,n_2,n_3)$ is a triple of positive
integers and  $s \in \{1,-1\}$ denotes a sign.  Let $E_{\vn,s}^{(0)}$
denote the corresponding umperturbed energies.  The first order
correction to the energy levels due to the Lorentz-violating terms are
found using standard  degenerate perturbation theory as: 
\begin{equation}
\langle \psi_{\vn,s}| H-p^2/2m |  \psi_{\vn,s} \rangle
= \frac{\pi^2 \hbar^2}{2 m L^2} \left(A n^2 + \sum_i F_{ii}n_i^2 
+ s |\gn|\right),
\label{EA1}
\end{equation}
where the vector $\gn$ is defined with components
\begin{equation}
(\gn)_j \equiv \frac{2 m L^2}{\pi^2 \hbar^2}B_j + \sum_i G_{iij}n_i^2
\quad .
\end{equation}

Using the perturbed energy expression (\ref{EA1}) and standard
approximations, the partition function is 
\begin{eqnarray}
Z^{(1)}  &  \simeq & 2e^{-\beta A}n_Q V \left(1 - \haf \tracce(F)\right)
\quad ,
\label{ZA}
\end{eqnarray}
where $V$ is the volume of the box,  $n_Q = 
({m}/{2\beta \pi \hbar^2})^{{3}/{2}}$ is the
quantum concentration, and
$\tracce(F) = \sum F_{ii}.$  It follows that only the $A$ term
corrects the energy and corresponds to a constant shift in all of the
energy levels. 

The correction to the partition function due to the $F$ term can 
be incorporated into an effective mass for the fermion 
\begin{equation}
m^* = \left(1 - \third \tracce(F)\right) m
\quad .
\label{effmass}
\end{equation}

The expectation value of the spin can be calculated similarly:
\begin{eqnarray}
\la {\bf s}^{(1)} \ra  & \simeq & - \beta {\bf B} - \frac{1}{2} \tracce(\bf G) 
 \quad ,
\label{SP1G} 
\end{eqnarray}
where the vector $\tracce({\bf G})$ is defined by
$(\tracce({\bf G}))_k \equiv \sum_i G_{iik}.$
\section{Classical gas}
The grand partition function for the classical gas 
system can be written in terms of the single-particle partition
function and a chemical potential.   This gives expressions for the
expected particle number and energy.  From these expressions it follows
that there is no change in the ideal gas law.   

It is possible to solve for the chemical potential $\mu^{(C)}$ 
\begin{eqnarray}
\mu^{(C)} & = & -kT\left( \ln\left(\frac{2n_Q}{n^{(C)}}\right) -
\haf\tracce( F) \right) 
\quad ,
\label{chem1} 
\end{eqnarray}
where $n_Q$ is the quantum concentration and $n^{(C)} \equiv {\la
N^{(C)} \ra}/{V}$ is the concentration of the classical gas.
One can also solve for the entropy, and there is a modification of the
Sackur-Tetrode equation: 
\begin{eqnarray}
S^{(C)} & = &  \la N^{(C)} \ra k\left[ \frac{5}{2} -
\frac 1 2 \tracce (F) +  \ln\left(\frac{2n_Q}{n^{(C)}}\right)
\right]. \label{ST1} 
\end{eqnarray}

Finally, the expectation of the spin is 
\begin{equation}
\la {\bf s}^{(C)} \ra = - \la N^{(C)} \ra \left[ \beta {\bf B} 
+ \haf \tracce(\bf G) \right]
\quad .
\end{equation}
\section{Quantum Gas - Fermions}
With notation from previous sections, and using zero subscripts to
represent unperturbed quantities, the partition function for the grand
canonical ensemble associated to a Fermi gas is given by 
\begin{eqnarray}
\ln\left(Z^{(Q)}_0(\alpha_0)\right)  & = & \frac{2}{\lambda^3} V
 f_{\frac{5}{2}}(e^{-\alpha_0})  
 \quad ,
 \label{Qintegral}
\end{eqnarray}
where $\lambda = {h}/{(2\pi mkT)^{\frac{1}{2}}}$ is the thermal
wavelength and $f_{\nu}(e^{-\alpha})$ is the appropriate Fermi-Dirac
integral.\cite{pathria}  As in the classical case, first order
corrections to the partition function occur only for Lorentz
violating terms of type $F.$  The corresponding partition function can
be written as  
\begin{eqnarray}
\ln\left(Z^{(Q)}(\alpha)\right) & \simeq & \left( 1 - \haf
 \tracce(F)\right) \ln\left(Z^{(Q)}_0(\alpha) \right)~. \label{QZ2}
\end{eqnarray}
As above, the perturbation can be absorbed as an effective
mass. Calculations then give 
\begin{eqnarray}
\la N^{(Q)}(\alpha) \ra & = &  \left( 1 -  \haf
\tracce(F)\right) \frac{2}{\lambda^3} f_{\frac{3}{2}}(e^{-\alpha})
\quad ,
\label{QN2}\\ 
\la E^{(Q)}(\alpha) \ra & = &   \frac{3}{2} \la N^{(Q)}(\alpha)
\ra kT\frac{f_{\frac{5}{2}}(e^{-\alpha})}{
f_{\frac{3}{2}}(e^{-\alpha})}\quad ,
\label{QE2} 
\end{eqnarray}
as well as the ideal gas law
\begin{eqnarray}
\frac{P}{n^{(Q)} k T} & = &
\frac{f_{\frac{5}{2}}(e^{-\alpha})}{f_{\frac{3}{2}}(e^{-\alpha})}
\quad ,
\label{IGL2} 
\end{eqnarray}
where $n^{(Q)}$ is the concentration of the quantum gas.

Since the map $\alpha \to f_{\frac{3}{2}}(e^{-\alpha}) $ is
invertible, we have a formal expression for the chemical potential in
terms of the inverse ${\mathcal F}$
\begin{eqnarray}
\mu^{(Q)} & \simeq- k T {\mathcal F}\left(\frac{\lambda^3(m^*)
n^{(Q)}}{2}\right) \quad .
\label{chemQ1}
\end{eqnarray}

Equation (\ref{chemQ1}) can be used to obtain expressions for relevant
thermodynamic quantities; for example, the Fermi energy and associated
perturbations of the chemical potential at low temperature, low
temperature expressions for specific heat and entropy, and low
temperature perturbations of the  ideal gas law.\cite{cm3} 

The expectation value for the spin can be calculated in the quantum
regime using the fractional occupancies:
\begin{eqnarray}
\la {\bf s}^{(Q)} \ra & \simeq & -\la N^{(Q)} \ra 
\left[2 \frac{\beta}{\lambda^3}f_{\haf}(e^{-\alpha}) {\bf B}
+ \frac12 \tracce({\bf G})\right] .
 \label{spinG}
\end{eqnarray}
At low temperatures, the contribution from the $B$ term can be written as 
\begin{equation}
\la {\bf s}^{(Q)}_B \ra \simeq -  \la N^{(Q)} \ra 
\frac32 \frac{{\bf B}}{{\mathcal{E_F}}}
\left[ 1 - \frac{\pi^2}{12}\left( \frac{kT}{{\mathcal{E_F}}}\right)^2 \right] 
\quad 
\end{equation}
where ${\mathcal E_F}$ is the Fermi energy. 
\section{Quantum gas - Bosons}
It is possible to generate a model for a free spin-0 boson gas by combining
two fermions into a singlet representation of the spin group.
The resulting hamiltonian is given by 
\begin{eqnarray}
H & = & \frac{p^2}{2m} + A +  C_{j}\frac{p_j}{m} +
F_{jk}\frac{p_jp_k}{2m} 
\quad .
\label{H3}
\end{eqnarray}
Choosing the ground state energy to be zero, employing the notation of the
previous sections and making the standard approximation, the
associated grand partition function for the unperturbed case can be
written as
\begin{equation}
\ln(Z^{(QB)}(\alpha_0)) =  \frac{1}{\lambda^3}Vg_{\frac{5}{2}}
(e^{-\alpha_0}) - \ln(1-e^{-\alpha_0})  
\quad ,
\end{equation}
where $\lambda $ is the thermal wavelength, 
$g_{\frac{5}{2}}(e^{-\alpha})$ is the appropriate Bose-Einstein 
integral, and the ground state has been separated out.  The only nontrivial
leading order perturbation in (\ref{H3}) arises from the $F$ term.  A
calculation which follows that done for the case of fermions gives  
\begin{eqnarray}
\ln\left(Z^{(QB)}(\alpha)\right) & \simeq & \left( 1 - \haf
 \tracce(F)\right) \frac{1}{\lambda^3}Vg_{\frac{5}{2}} (e^{-\alpha}) \nonumber \\
 & & -
\ln(1-e^{-\alpha}) \quad . 
\label{QZB2}
\end{eqnarray}
It follows that for the perturbed case we have  
\begin{eqnarray}
\la N^{(QB)}(\alpha) \ra - \la N_{G0} \ra & = &  \left( 1 -  \haf
\tracce(F)\right) \frac{1}{\lambda^3} g_{\frac{3}{2}}(e^{-\alpha}) ~~,
\nonumber \\ 
\la E^{(QB)}(\alpha) \ra & = &   \frac{3}{2} kT \frac{V}{\lambda^3}
g_{\frac{5}{2}}(e^{-\alpha}) \quad .
\label{QEB2}  
\end{eqnarray}
As in the Fermi case, the chemical potential can be expressed 
as a function of the number of particles in excited states.  Because
only $\tracce(F)$ enters into the grand partition function, it is
possible to use the concept of effective mass to absorb the effect as
before.  Standard results of Bose-Einstein condensation therefore hold
in a given laboratory frame.  
\section{Conclusion}
Using Jaynesian formalism, we have developed a framework for
statistical mechanics in the presence of symmetry violation which
parallels the conventional case.  We find that the laws of
thermodynamics are the same as in the conventional case, with specific
expectation values of thermodynamic quantities modified by the
Lorentz-violating terms.

For an ideal gas in the absence of any external applied fields,
expectation values for scalar thermodynamic quantities such as energy
and particle number were unaltered except for an overall scaling
factor $\tracce (F)$.  This correction can be incorporated into theory as
an effective mass $m^* = (1 - \third \tracce (F)) m$ in the
hamiltonian, although the effective mass defined in this way depends
on the observer's Lorentz frame.   

Focussing on spin, we find nontrivial changes in the net spin
expectation value arising from the terms that couple to the spin. 
The pure-spin coupling $B_j$ mimics a constant background magnetic  
field and induces a corresponding magnetic moment per unit volume in
the gas.  The derivative-spin coupling $G_{ijk}$ generates a
fundamentally new type of effect that induces a
temperature-independent polarization in the classical gas that is
proportional to $\tracce({\bf G})$. 


\end{document}